\documentclass[twocolumn,floats,aps,showpacs]{revtex4}
\usepackage{times,amsmath,graphics,psfrag,latexsym}

\begin{document}
\title{Heavy-ion irradiation of UBe$_{13}$ superconductors}
\author{H.A. Radovan$^1$, E. Behne$^1$, R.J. Zieve$^1$, J.S. Kim$^2$, G.R. Stewart$^2$, 
W.-K. Kwok$^3$, and R.D. Field$^4$},
\affiliation{${}^1$Physics Department, University of California at Davis\\
${}^2$Physics Department, University of Florida\\
${}^3$Argonne National Laboratory, Division of Materials Science\\
${}^4$Los Alamos National Laboratory, Division of Materials Science
and Technology}
\begin{abstract}
We irradiate the heavy fermion superconductors (U,Th)Be$_{13}$ with high-energy
heavy ions. Damage from the ions affects both heat capacity and magnetization
measurements, although much less dramatically than in other superconductors.  
From these data and from direct imaging, we conclude that the irradiation does
not create the amorphous columnar defects observed in
high-temperature superconductors and other materials.  We also find
that the damage suppresses the two superconducting transitions of
U$_{0.97}$Th$_{0.03}$Be$_{13}$ by comparable amounts, unlike their
response to other types of defects.  
\end{abstract} 
\maketitle

\section{Introduction}

Microscopic lattice disorder can have interesting effects, particularly on
unconventional superconductors. 
Sensitivity of the critical temperature $T_c$
to non-magnetic scatterers serves as evidence of non-$s$-wave electron-pairing.
Defects can be introduced during sample fabrication, or induced afterwards by
irradiation. The latter possibility encompasses both point defects created by
light ions and other particles, and amorphous columns created by high-energy
heavy ions. These columnar defects, which have been heavily studied in
high-temperature superconductors (HTS), serve as highly efficient pinning
centers for vortices and alter a sample's response to a magnetic field.

Little work has been done on another group of unconventional
superconductors, the heavy fermion (HF) systems.  The two orders of
magnitude difference in the critical temperatures of HTS and HF
superconductors changes the scale for thermal effects.  
Thermal fluctuations which complicate HTS behavior should be absent
in HF.

The spectacular response of one heavy fermion material, UBe$_{13}$, to thorium
doping makes it a good candidate for investigating other forms of lattice
disorder. $T_c(x)$ is not monotonic for U$_{1-x}$Th$_x$Be$_{13}$, and a second
thermodynamic phase transition appears for $0.019 < x < 0.043$ 
\cite{2trans,muSR}.  No other dopant has such effects.  The Be as
well as the U is sensitive to substitution details. The superconducting
transition is suppressed below 0.015 K for UBe$_{12.94}$Cu$_{0.06}$
\cite{Andraka}, while the same amount of B has virtually no effect
\cite{boron}.

Multiple superconducting phases indicate an unconventional order parameter.  A
second HF material, UPt$_3$, also exhibits a double transition \cite{Fisher}. 
Furthermore, the lower phases of both UPt$_3$ and U$_{1-x}$Th$_x$Be$_{13}$ have
a significantly reduced vortex relaxation rate, while in pure UBe$_{13}$ the
relaxation rate is linear in temperature \cite{Mota}.  This may be evidence of
time reversal symmetry breaking in the lower phase. If time reversal symmetry
is broken, spontaneous magnetic fields can appear.  Boundaries between
different magnetic domains can obstruct vortex motion, leading to the observed
drop in relaxation rate. Vortex motion is also reduced in
U$_{1-x}$Th$_x$Be$_{13}$ for zero-field-cooling as compared to the
field-cooling case \cite{hole}.  Domain wall pinning may explain this effect as
well, since cooling in a field should lead to larger domains and fewer domain
walls.   The atypical vortex behavior and the distinction between the two
superconducting phases suggests that vortices may also have interesting
interactions with other types of pin sites, such as columnar defects.

Here we investigate heavy-ion irradiation of (U,Th)Be$_{13}$. We show that
calculations predict columnar tracks to form.  However, heat
capacity measurements, imaging, and magnetization measurements suggest that
such tracks are not actually present.  We also discuss how heavy-ion damage
affects the different superconducting phases.  

\section{Irradiation Damage}

The thermal spike model explains damage creation in both elements and
alloys \cite{Wang,Ghidini,Nozieres}.  In this model, the heavy ion loses energy
primarily to the electrons, which in turn excite phonons.  The electron-phonon
coupling strength $g$ determines how fast the energy is transfered to the
lattice, while the thermal conductivities $\kappa_e$ and $\kappa_p$ of
electrons and phonons govern how the energy spreads through the sample.  The
coupled differential equations 
\begin{equation*}
C_e\frac{\partial T_e}{\partial t}=\frac{\partial}{\partial r}
(\kappa_e\frac{\partial T_e}{\partial r})+\frac{\kappa_e}{r}
\frac{\partial T_e}{\partial r}-g(T_e-T_p)+A(r,t) 
\label{e:electrons}
\end{equation*}
\begin{equation*}
C_p\frac{\partial T_p}{\partial t}=\frac{\partial}{\partial r}
(\kappa_p\frac{\partial T_p}{\partial r})+\frac{\kappa_p}{r}
\frac{\partial T_p}{\partial r}+g(T_e-T_p)
\label{e:phonons}
\end{equation*}
describe the energy transfer. Here $\kappa_e$ and the 
specific heat $C_e$ of the electron system are functions of the electron
temperature $T_e$, while $\kappa_p$ and the lattice specific heat $C_p$ 
depend on the phonon temperature $T_p$.  
$S_e = -\frac{dE}{dx}|_e$ is the energy loss from the heavy ion to the
electrons in the target.
As $S_e$ increases, a threshold
value allows formation of amorphous tracks.  With further increase the track
radius grows.  For the radial distribution of the original ion energy loss we
use
\begin{equation*}
\hspace{-1in}
A(r,t)\propto\frac{1}{r}[\frac{(1-\frac{r+\theta}{r_{max}+\theta})^{0.927}}
{r+\theta}]
\end{equation*}
\begin{equation*}
\hspace{.2in}\times[1+\frac{K(r-L)}{M}e^{-(r-L)/M}]e^{-(t-t_o)^2/2t_o^2},
\label{e:ionloss}
\end{equation*}
as in \cite{ionloss}.  Here $K=19\beta^{1/3}$, $L=0.1$ nm,  $M=(1.5+0.5\beta)$
nm, and $\beta$ is the ion velocity relative to the speed of light.  The
maximum range $r_{max}=6\times 10^{-6} (\frac{2mc^2\beta^2}
{1-\beta^2})^{1.079}$ cm, where $m$ is the electron mass in grams  and $c$ the
speed of light in cm/s.  The time $t_o$ is of order $10^{-15}$ s and
$\theta=9.84\times 10^{-9}$ cm. 
The normalization of $A(r,t)$ is chosen
so that the total energy transfer to the electrons equals $S_e$.
This formula, particularly the term involving $K$, $L$, and $M$, comes from 
fits to observed radiation distributions in water \cite{ionloss}.
Thermal spike calculations using these equations successfully predict the existence
and even radius  of columnar defects in a variety of pure metals and alloys
\cite{Wang}.

\begin{table}[t]
\caption{Parameters for thermal spike calculations in UBe$_{13}$.}
\label{t:thspike}
\renewcommand{\tabcolsep}{1pc} 
\renewcommand{\arraystretch}{1.2} 
\begin{tabular}{@{}lll}
\hline
Debye temperature, $\Theta_D$ & 620 K & \cite{Stewartrev} \\
Melting temperature, $T_m$ & 2273 K & \cite{deltaH}\\
Resistivity, $\rho$ & 110 $\mu\Omega$-cm at 300 K& \cite{rho}\\
Atomic density, $n_a$& $1\times 10^{23}$/cm$^3$ & \cite{xtal}\\
Valence, $z$ & 2.1 & \\
$C_e$ & $4.3\times 10^3$ erg/cm$^3$ K & \\
$C_p$ & $4.1\times 10^3$ erg/cm$^3$ K & \\
$\kappa_e$ & $2.2\times 10^3T$ erg/cm K & \\
$\kappa_p$ & $9.9\times 10^7/T$ erg/cm K & \cite{Aliev}\\
\hline
\end{tabular}\\[2pt]
\end{table}

\begin{figure}[thb]
\begin{center}
\psfrag{dEdx (keV/nm)}{\scalebox{1.5}{$S_e$ (keV/nm)}}
\psfrag{Se (keV/nm)}{\scalebox{1.8}{$S_e$ (keV/nm)}}
\scalebox{0.4}{\includegraphics{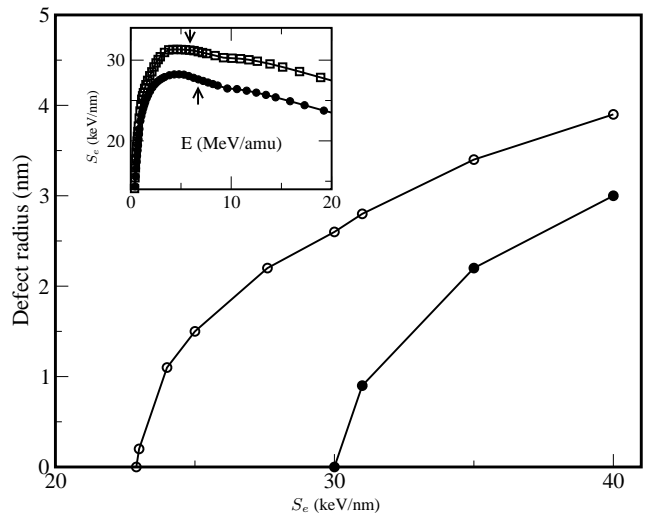}}
\caption{\small Predicted defect radius in UBe$_{13}$.  The two curves
use heat of fusion $2\times 10^{10}$ erg/cm$^3$ ($\bullet$) and
$4\times 10^{10}$ erg/cm$^3$ ($\circ$).
Inset: SRIM simulation of electronic energy loss for ${}^{238}$U$^{67+}$
($\Box$) and ${}^{208}\mbox{Pb}^{56+}$ ($\bullet$) ions in UBe$_{13}$.
The arrows indicate the energy of the beams used for our irradiations.}
\label{f:radius}
\end{center}
\end{figure}

Numerical solution of these equations requires knowledge of various parameters
of the material.  Table \ref{t:thspike} shows the values we use. Since most of
the simulation is at high temperature, we use high-temperature limits in
several cases.  The electronic and lattice specific heats are $C_e=1.5k_Bn_e$
and $C_p=3k_Bn_a$.  The density of atoms in the lattice is
$n_a$, while the electron density is $n_e=zn_a$ with $z$ the average
valence of the atoms composing the material.  For the electronic thermal 
conductivity $\kappa_e$ we use
the Wiedemann-Franz law.  Although in most metals the  high-temperature
electrical resistivity is proportional to temperature, in UBe$_{13}$ the
resistivity is nearly constant, so $\kappa_e$ is itself proportional to
temperature \cite{rho}.  For the lattice thermal conductivity $\kappa_p$ we
expect a $1/T$ temperature dependence and obtain the prefactor from the 90 K
value \cite{Aliev}. 

The heat of fusion of a compound equals the latent heat of its component
elements, plus the difference between the liquid state mixing enthalpy and the
solid enthalpy of formation. Typical values are $10^{10}$ to $10^{11}$
erg/cm$^3$.  In this case the U and Be latent heats contribute $2\times
10^{10}$ erg/cm$^3$. The heat of formation at room temperature is $-2\times
10^{10}$ erg/cm$^3$ \cite{deltaH}, so $4\times 10^{10}$ erg/cm$^3$ is an upper
bound on the heat of fusion.  Although the mixing enthalpy is unknown, $2\times
10^{10}$ erg/cm$^3$ is a more reasonable estimate for the heat of fusion.

Finally, for the electron-phonon coupling we use \cite{Wang}
$$g=\frac{\pi^4k_B^4n_e^2\Theta_D^2\rho}{18L\hbar^2(6\pi^2n_a)^{2/3}}
\frac{1}{T_e}.$$ Here $L=0.245$
erg-$\Omega$/K is the Lorentz number.

Figure \ref{f:radius} shows the defect radius obtained by solving the thermal
spike equations numerically.  The values are comparable to those of Ce and Y
alloys \cite{Ghidini}, which have similar resistivity, Debye temperature, and
heat of formation to UBe$_{13}$.  The high atomic density of UBe$_{13}$
increases its sensitivity to irradiation, while the unusual temperature
independence of the resistivity decreases the sensitivity.

\begin{table*}[t]
\caption{Effects of irradiation on transitions in (U,Th)Be$_{13}$.}
\label{t:tc}
\newcommand{\m}{\hphantom{$-$}}
\newcommand{\cc}[1]{\multicolumn{1}{c}{#1}}
\renewcommand{\tabcolsep}{1pc} 
\renewcommand{\arraystretch}{1.2} 
\begin{tabular}{@{}lccccc}
\hline
& $T_{c}$ (0 T) & $T_c$ (3 T) & $T_c$ (5 T) & $T_c$ (7 T) & \% change at 7 T \\
\hline
UBe$_{13}$ & 770 &  & 750 & 715$^*$ & -7.1 \\
U$_{0.97}$Th$_{0.03}$Be$_{13}$ ($T_{c1}$) & 625 & 620$^*$ & 613 & 530 & -15.2 \\
U$_{0.97}$Th$_{0.03}$Be$_{13}$ ($T_{c2}$) & 412 &  & 400 & 345 & -16.3 \\
\hline
\end{tabular}\\[2pt]
Transition temperatures marked by (*) come from magnetization measurements, the
remainder from heat capacity.
\end{table*}

\section{Sample Preparation and Imaging}

Our samples are polycrystalline UBe$_{13}$ and U$_{0.97}$Th$_{0.03}$Be$_{13}$
pieces with a grain size of about 100 $\mu$m, prepared by arc melting in an argon
atmosphere and subsequent annealing at 1400$^\circ$ C in a beryllium atmosphere
for 1000 hours \cite{anneal}.   Before irradiation, samples are thinned to $25\
\mu\mbox{m}$ to prevent ion implantation, with a typical lateral dimension of 1
mm. The irradiation is performed at ATLAS, Argonne National Laboratory with 1.4
GeV ${}^{238}\mbox{U}^{67+}$ and ${}^{208}\mbox{Pb}^{56+}$ ions.  The ion
velocities are 5.9 MeV/amu and 6.7 MeV/amu, respectively.  The matching fields
$B_\Phi$, at which the vortex density and track density are equal, range from 0.1
T and 7 T, corresponding to typical lateral track separations from 155 to 18.5
nm.  The ${}^{208}\mbox{Pb}^{56+}$ ions were used for the 3 T and 7 T samples,
while the ${}^{238}\mbox{U}^{67+}$ were used for the remaining samples. The
irradiation takes up to two hours, depending on dosage.  Because of the very small
probability that two ions will reach the sample at nearly the same time and
position, we assume that each ion acts independently in damaging the sample.

The inset of Figure \ref{f:radius} shows the electronic energy loss $S_e$
as a function of the incident beam energy, obtained from a
Stopping and  Range of Ions in Matter (SRIM-98 code \cite{SRIM}) simulation. 
The beam energy used for the irradiation  is near the maximum and gives
$S_e=31.2$ keV/nm for uranium, $S_e=27.6$ keV/nm for lead.  The 
corresponding nuclear energy losses are negligible, 0.05 keV/nm and 0.04 keV/nm,
respectively.  The predicted stopping distances are 56 $\mu$m and 60 $\mu$m,
more than twice the sample thickness.  From our thermal spike calculation,
the irradiation should produce defects with radius 2.8 nm (uranium) or
2.2 nm (lead).

After irradiation, both the 2.0 T UBe$_{13}$ and 7.0 T 
U$_{0.97}$Th$_{0.03}$Be$_{13}$ crystals were further thinned and viewed in both
a Philips CM-30 transmission electron microscope (TEM) and a JEOL 3000F high
resolution transmission electron microscope.  The planes used for the images on
the JEOL machine limited the resolution to about 0.5 nm.  At $B_\Phi=$2.0 T the
defects should be about 32 nm apart. The total viewing area of about 73,000
nm$^2$ should contain about 70 defects but in fact shows no evidence of
amorphous columns.  For the 7.0 T sample, the imaging area of over 54,000
nm$^2$ should have nearly 200 defects.  We again find no damage of radius
greater than 2 nm, and only six candidates with radius from 1 nm to 2 nm.  Of
these, closer examination shows that three are edge dislocations.  The
remaining candidates could also be dislocations, viewed from angles that do not
clearly show a lattice plane ending. We find no compelling reason to identify
them with columnar tracks, since they do not match the expected defects either
in size or in density.  We conclude that columnar defects are either absent in
our sample or far smaller than anticipated from our thermal spike model
calculations.

We emphasize that the microscopes we used {\em can} detect amorphous regions of
the expected size.  A transmission electron microscope successfully imaged
columnar defects of radius 5 nm in UO$_2$ \cite{imageTEM}.  High resolution TEM
has detected columnar defects with radius down to 1 nm, for example in tin
oxide \cite{imageHREMtin} and Bi$_2$Sr$_2$CaCu$_2$O$_{8+\delta}$
\cite{imageHREM}.  

\begin{figure}[thb]
\begin{center}
\psfrag{C/T (J/moleK2)}{\scalebox{1.6}{$C/T$ (J/mole K$^2$)}}
\scalebox{0.5}{\includegraphics{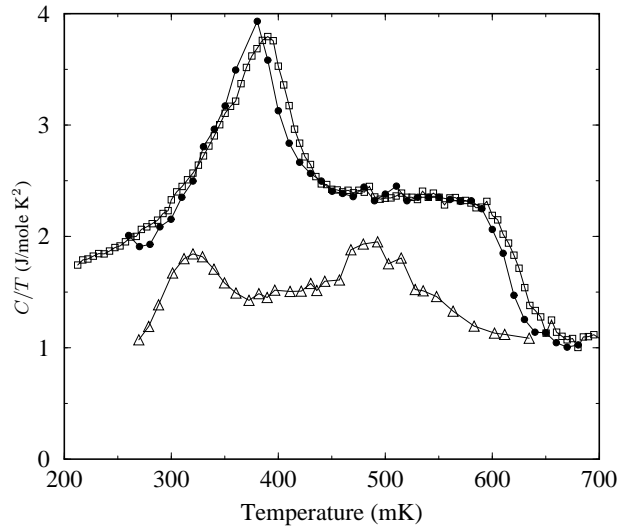}}
\caption{\small Zero-field heat capacity of unirradiated ($\Box$),
5 T irradiated ($\bullet$), and 7 T irradiated ($\Delta$)
U$_{0.97}$Th$_{0.03}$Be$_{13}$.}
\label{f:cvst}
\end{center}
\end{figure}

\section{Heat Capacity Measurements}

We measure heat capacity on a dilution refrigerator at temperatures down to 100
mK and magnetic fields up to 8 Tesla. We use a relaxation method with a RuO$_2$
thin film thermometer, a [50:50] AuCr thin film heater, and a fine copper wire
as a heat link. 

Figure \ref{f:cvst} shows $C/T$ for the unirradiated, 5 T, and 7 T
U$_{0.97}$Th$_{0.03}$Be$_{13}$ samples.  Since we do not know the precise
sample sizes, we use a multiplicative constant to set the normal state value of
$C/T$ to  1.1 $\mbox{J/mole K}^2$ for each curve.   We do not adjust the
measured heat capacity for any contribution from the thermometer, heater, or
mount; but comparing the curves shown to previous heat capacity measurements on
bulk samples suggests that background effects are less than 30\% of our signal.
In any case, the background should not affect our determination of the $T_c$
suppression or transition width.

Irradiation does not measurably change $T_c$ up to $B_\Phi=2$ T. As shown in Figure
\ref{f:cvst} for U$_{0.97}$Th$_{0.03}$Be$_{13}$, $C/T$ changes shape only slightly
even at a defect density of 5 T, broadening by several mK \cite{Tcred}.  At the
higher defect density of $B_\Phi=7$ T, the transitions do change shape. The upper
transition widens by roughly a factor of two, and both decrease significantly in
amplitude.  Here width is defined by identifying the change in $C/T$ across the
transition and using the temperature difference between the 10\% and 90\% values of
the heat capacity. Irradiation has a similar effect on pure UBe$_{13}$, reducing
$T_c$ and slightly broadening the transition, while retaining the shape of the $C/T$
curve up to 5 T. Table \ref{t:tc} summarizes the effects of irradiation on 
transition temperatures for both compounds.  The marked transitions come from
magnetization measurements rather than heat capacity, as will be discussed later.  

Since the ${}^{208}\mbox{Pb}^{56+}$ ions used for irradiation have less energy
loss than the ${}^{238}\mbox{U}^{67+}$ ions, they produce smaller defects in a
thermal spike calculation.  Depending on the value of the heat of formation, the
uranium atoms might even produce defects while the lead atoms do not.  We note
here that the $T_c$ depression for the 3 T sample is consistent with the others.
The 7 T samples have, if anything, a larger change in $T_c$ than expected from
the lower irradiation dosages.  Thus it appears that the
${}^{208}\mbox{Pb}^{56+}$ ions do as much damage as the
${}^{238}\mbox{U}^{67+}$ ions, despite their smaller energy loss.

It is tempting to take the nearly identical percentage suppression of the two
thoriated transitions as evidence that both superconducting phases have the
same pairing symmetry.  However, the lower transition is between two
superconducting phases, rather than between a superconducting and a
normal phase.  The usual arguments on how impurities affect the transition
do not apply.  We do note that the heavy-ion irradiation produces a
different response from either non-magnetic or magnetic substitutional
impurities \cite{Scheidt}.  Non-magnetic (La) impurities depress the upper
transition but leave the lower transition unchanged, while magnetic (Gd)
impurities suppress the lower transition more than twice as strongly
as the upper one \cite{Scheidt}.  These contrasting behaviors of the
two transitions for different types of defects emphasize the extreme
sensitivity of the (U,Th)Be$_{13}$ system to changes in electronic
structure.

The $T_c$ suppression proves that the irradiation did create some damage
within the crystals.  However, the small amount of suppression, like
the lack of visible columns in the imaging, suggests that the damage
is {\em not} in the form of amorphous columns.  For comparison,  U ions
at a 2 T matching field reduce $T_c$ of YBa$_2$Cu$_3$O$_{7-x}$ by 5.3\%
\cite{UinHTS}, more than we find at 5 T.  Although in principle our low
$T_c$ reduction could occur from insensitivity to defects, rather than
an absence of defects, HF superconductors are {\em more} sensitive
than other superconductors to both neutron and light-ion irradiation
\cite{Andraka,lightion}.  Even our $B_\phi=7$ Tesla irradiation affects
$T_c$ far less than the point defect work \cite{Andraka}.

\begin{figure}[thb]
\begin{center}
\psfrag{DM (G)}{\scalebox{1.5}{$\Delta M$ (gauss)}}
\psfrag{C/T (J/mol K2)}{\scalebox{1.5}{$C/T$ (J/mol K$^2$)}}
\psfrag{Tc1}{\scalebox{1.4}{$T_{c1}$}}
\psfrag{Tc2}{\scalebox{1.4}{$T_{c2}$}}
\scalebox{0.65}{\includegraphics{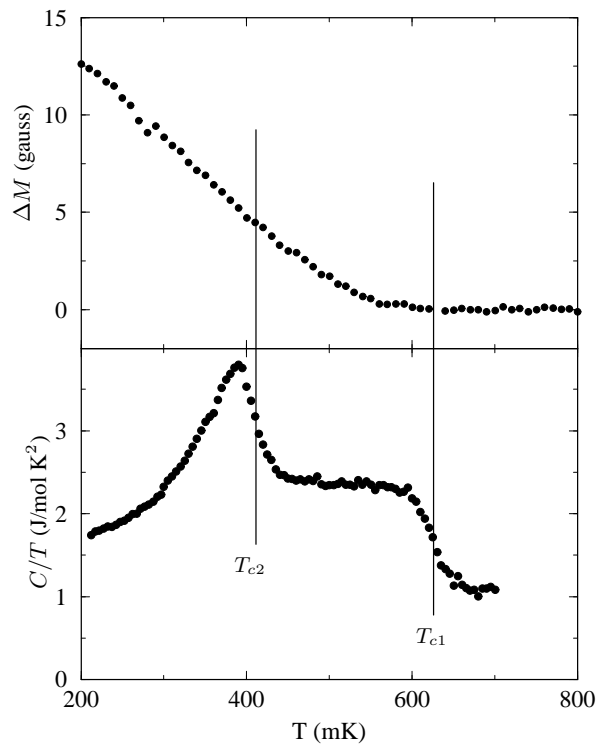}}
\caption{\small Comparison of magnetization and heat capacity for 
unirradiated U$_{0.97}$Th$_{0.03}$Be$_{13}$.  The vertical lines mark
the two transitions.}
\label{f:magandc}
\end{center}
\end{figure} 

\section{Magnetization Measurements}

The motion of superconducting vortices is particularly sensitive to
defects.  Columnar tracks strongly pin vortices, even when
the tracks are misaligned with the magnetic field by up to 40$^\circ$ \cite{tilt}.
Our imaging work could miss incomplete or misaligned columns which
would be detected in a sample's magnetic behavior. 

For these measurements we use bismuth Hall probes with active area
$10\times 15\ \mu\mbox{m}^2$.  We place the sample flat atop the
substrate for the Hall  probe.  The probe measures total field at a spot
near the surface of the sample, from which we extract the local
magnetization of the superconductor. We typically ramp the applied
magnetic field at 15 gauss per minute, and take hysteresis loops large
enough to allow for full penetration.  The width of the hysteresis loop,
$\Delta M$, is proportional to the critical current $j_c$; and we measure
both its temperature and its field dependence.

Figure \ref{f:magandc} compares hysteresis loop width and heat capacity for
a sample.  As $T$ increases, the hysteresis loop width decreases, closing
at the same superconducting $T_c$ measured with heat capacity.  For the 3 T
thoriated and 7 T pure samples, the $T_c$ values in Table \ref{t:tc} are
determined from magnetization rather than heat capacity data.

We note that $\Delta M(T)$ in Figure \ref{f:magandc} shows no feature at the
lower transition. By contrast, magnetic relaxation measurements show a sharp
drop in the vortex creep rate at the lower transition \cite{Mota}.  UPt$_3$,
the other HF compound with two transitions, shows the same behavior: a drop in
the relaxation rate at the lower transition \cite{Mota}, but no feature in the
magnetization itself \cite{Emmin}. One explanation of the difference between
the two measurements is that the mechanism of vortex motion changes at the
lower transition, while the pinning strength itself undergoes no sharp change. 
However, another possibility lies in the magnetic field history of the
relaxation measurements. Vortex creep is only suppressed 
when the sample is exposed to a sufficiently large magnetic field,
typically 1 kOe, before having the field reduced to zero for the
relaxation measurements \cite{Mota2}.  Our hysteresis loops, with maximum
width of 200 Oe, may simply be in a different regime where the vortex
creep shows no signature at the lower transition.

The magnetization data confirm that columnar defects are either absent or far smaller
than expected. Increased pinning from the columns would appear as wider hysteresis
loops, but our loop widths are comparable for unirradiated and irradiated samples. 
In other superconductors columnar defects produce large effects from commensurability
near integral multiples of $B_\Phi$, but our hysteresis loops have no features 
whatsoever in $B$. One possibility is that any defects are too much smaller than 100
\AA, the superconducting coherence length \cite{rho}, to pin vortices.

\begin{figure}[thb]
\begin{center}
\psfrag{dM (g)}{\scalebox{2}{$\Delta M$ (gauss)}}
\psfrag{1-T/Tc}{\scalebox{2}{$1-\frac{T}{T_c}$}}
\scalebox{0.5}{\includegraphics{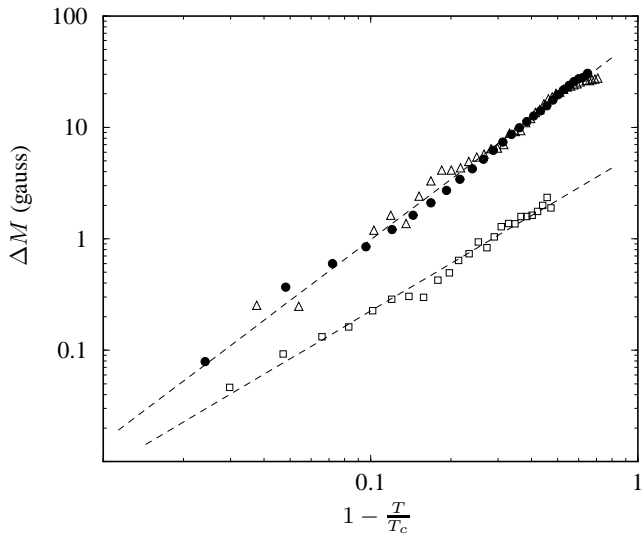}}
\caption{\small Hysteresis loop width vs. reduced temperature for three 
thoriated samples: unirradiated ($\bullet$), and irradiated with matching
fields of 5 T ($\bigtriangleup$) and 7 T ($\Box$).}
\label{f:loglog}
\end{center}
\end{figure}

Our magnetization measurements do show evidence of defect creation.
Figure \ref{f:loglog} shows loop width versus normalized temperature for
(U,Th)Be$_{13}$ crystals of several irradiation doses, taken at zero nominal
field.  The width follows a power law $\Delta M(T) =
\Delta M(0)(1-\frac{T}{T_c})^\alpha$ over the entire temperature range measured,
approximately 0.3$T_c$ to $T_c$.  For the unirradiated and the 5 T samples
$\alpha = 1.8$, while for the 7 T specimen $\alpha$ is reduced to 1.4.  
Although the magnitude agreement between the 5 T and unirradiated samples is
partly coincidental, the 7 T sample has significantly the smallest $\Delta M$
observed.  Both the magnitude and slope change are probably due to the
damage.  As with the stronger $T_c$ suppression in the 7 T sample mentioned
earlier, this hints that the 5 T irradiation is a threshold for strong damage
in our (U,Th)Be$_{13}$ crystals.  

Models for networks of superconducting grains give $\alpha=1$ for
superconductor-insulator-superconductor (SIS) tunnel junctions \cite{AB} and
$\alpha=2$ for superconductor-normal-superconductor (SNS) junctions
\cite{deGennes}.
An intermediate power law for $j_c(T)$, with $\alpha$ near 1.5, has been
measured in various materials, including the HF material CeCu$_2$Si$_2$
\cite{CCSjc}.  The intermediate values may come from mixtures of SIS
and SNS junctions in the samples.  Similarly, the change in exponent
for our 7 T sample is consistent with irradiation damage changing the
characteristics of the grain network. We find the same power-law behavior 
in external magnetic fields up to $\mu_0H = 2$ kG, with the exponent 
independent of field.

\section{Conclusion}

We have irradiated (U,Th)Be$_{13}$ crystals with high-energy U and Pb ions at
fluences up to $B_{\Phi} = 7$ T.  The irradiation damage reduces both
superconducting transition temperatures of U$_{0.97}$Th$_{0.03}$Be$_{13}$ by
over 15\% for the 7 T bombardment, as well as lowering $j_c$ and reducing
its temperature dependence.  Unlike in previous work with substitutional
impurities \cite{Scheidt}, irradiation causes a similar $T_c$ reduction
for the two transitions of U$_{0.97}$Th$_{0.03}$Be$_{13}$.

Despite the measured $T_c$ reduction, transmission electron microscope and hysteresis
measurements suggest that the defects are not the amorphous tracks found in HTS. Our
calculations of the irradiation process predict columnar defect formation for $S_e$
down to 23 keV/nm.  Our samples are irradiated above the threshold value, at $S_e
\approx 30$ keV/nm.  The resulting defects should have diameter 5 to 6 nm, a defect
size comparable to the 10 nm superconducting coherence length.  The lack of pinning
by the defects and their absence from the TEM images means either that the actual
radius is much smaller or that only incomplete columns or point defects form.  We
favor the third option.  We rule out the first possibility because of the similarity
between ${}^{208}\mbox{Pb}^{56+}$ and ${}^{238}\mbox{U}^{67+}$ irradiation: if the
column radius were much smaller for ${}^{238}\mbox{U}^{67+}$ ions, then no tracks
should appear at all for ${}^{208}\mbox{Pb}^{56+}$ ions. The lack of signal in
the magnetization indicates that the sample does not have even incomplete
columns, which would still pin vortices effectively.  Since the reductions of
$T_c$ and $j_c$ show that some damage is present, we assume we have instead
created point defects.

The conditions for creation of columnar defects are
important to understand, particularly as the use of such defects spreads from HTS to
other materials.
It is unclear whether the discrepancy between the experiment and the
thermal spike calculation comes from the values chosen for the parameters or from a
deeper problem with the model. The thermal spike model has successfully predicted
radiation damage both in other alloys \cite{Ghidini} and in elements, including
uranium and beryllium \cite{Wang}.
The simplest explanation would be incorrect parameters.  Although we use conservative
choices, such as assuming resistivity constant rather than linear in temperature,
our high-temperature extrapolations for $\rho$ or $\kappa_p$ may be incorrect.
Many-electron effects produce a variety of unusual low-temperature
effects in UBe$_{13}$ and other heavy fermion materials, and perhaps could
influence higher-temperature behavior as well.

We also show that the lower transition of U$_{0.97}$Th$_{0.03}$Be$_{13}$ is not
detectable in $j_c(T)$, despite the sharp change in magnetic relaxation rate
previously observed.  This may arise from differences in measurement procedure
between our work and the relaxation rate studies.  Another possibility is that
vortices in the two superconducting phases of U$_{0.97}$Th$_{0.03}$Be$_{13}$
differ only in their behavior away from the critical current $j_c$, which
magnetization measurements would not probe.

\section{Acknowledgements}

We would like to thank P.C. Canfield and J.L. Smith for fruitful
discussions.  This work was supported by NSF under DMR-9733898 (UCD) and
by DOE under W-31-109-ENG-38 (ANL), DE-FG05-86ER45268 (Florida),
and W-7405-ENG-36 (LANL).

\end{document}